\documentclass[aps,prl,reprint,superscriptaddress,showpacs]{revtex4-1}
\usepackage[latin9]{inputenc}
\usepackage{babel}
\usepackage{amsmath}
\usepackage{amssymb}
\usepackage{graphicx}
\usepackage[unicode=true,
 bookmarks=false,
 breaklinks=false,pdfborder={0 0 1},colorlinks=true,linkcolor=red,citecolor=blue]
 {hyperref}
\makeatletter
\@ifundefined{textcolor}{}
{%
 \definecolor{BLACK}{gray}{0}
 \definecolor{WHITE}{gray}{1} 
 \definecolor{RED}{rgb}{1,0,0}
 \definecolor{GREEN}{rgb}{0,1,0}
 \definecolor{BLUE}{rgb}{0,0,1}
 \definecolor{CYAN}{cmyk}{1,0,0,0}
 \definecolor{MAGENTA}{cmyk}{0,1,0,0}
 \definecolor{YELLOW}{cmyk}{0,0,1,0}
}

\usepackage{amsthm}
\usepackage{amsfonts}
\usepackage{bbm}
\usepackage{epsfig}
\usepackage{times}
\usepackage{babel}
\usepackage{color}
\usepackage{framed}
\usepackage{changes}
\usepackage{float}
\usepackage{array}
\usepackage{comment}
\makeatletter

\providecommand{\tabularnewline}{\\}

\@ifundefined{textcolor}{}
{%
 \definecolor{BLACK}{gray}{0}
 \definecolor{WHITE}{gray}{1}
 \definecolor{RED}{rgb}{1,0,0}
 \definecolor{GREEN}{rgb}{0,1,0}
 \definecolor{BLUE}{rgb}{0,0,1}
 \definecolor{CYAN}{cmyk}{1,0,0,0}
 \definecolor{MAGENTA}{cmyk}{0,1,0,0}
 \definecolor{YELLOW}{cmyk}{0,0,1,0}
}

\usepackage{amsthm}
\usepackage{amsfonts}
\usepackage{bbm}
\usepackage{epsfig}
\usepackage{times}
\usepackage{babel}
\usepackage{color}
\usepackage{framed}
\usepackage{changes}
\usepackage{float}
\usepackage{array}

\makeatother

\begin{document}
\title{Dynamical transitions in scalarization and descalarization through black hole accretion}
\author{Cheng-Yong Zhang}
\email{zhangcy@email.jnu.edu.cn}

\address{\textit{Department of Physics and Siyuan Laboratory, Jinan University,
Guangzhou 510632, China }}
\author{Qian Chen}
\email{chenqian192@mails.ucas.ac.cn (corresponding author)}

\address{\textit{School of Physical Sciences, University of Chinese Academy
of Sciences, Beijing 100049, China }}
\author{Yunqi Liu}
\email{yunqiliu@yzu.edu.cn }

\address{\textit{Center for Gravitation and Cosmology, College of Physical
Science and Technology, Yangzhou University, Yangzhou 225009, China}}
\author{Wen-Kun Luo}
\email{luowk@stu2020.jnu.edu.cn}

\address{\textit{Department of Physics and Siyuan Laboratory, Jinan University,
Guangzhou 510632, China }}
\author{Yu Tian}
\email{ytian@ucas.ac.cn}

\address{\textit{School of Physical Sciences, University of Chinese Academy
of Sciences, Beijing 100049, China }}
\address{\textit{Institute of Theoretical Physics, Chinese Academy of Sciences,
Beijing 100190, China}}
\author{Bin Wang}
\email{wang\_b@sjtu.edu.cn (corresponding author)}

\address{\textit{Center for Gravitation and Cosmology, College of Physical
Science and Technology, Yangzhou University, Yangzhou 225009, China}}
\address{\textit{Shanghai Frontier Science Center for Gravitational Wave Detection, Shanghai Jiao Tong
University, Shanghai 200240, China}}
\begin{abstract}
%We present the first fully nonlinear study on the accretion of scalar fields onto a seed bald black hole in  anti-de Sitter spacetime in Einstein-Maxwell-scalar theory. We disclose intrinsic critical phenomena and explain why and how different scalarizations can happen. Further we construct a new physical mechanism to dynamically descalarize initial hair of an \red{isolated} scalarized black hole.  We reveal the first results on critical phenomena in descalarizations and compare with the corresponding properties in scalarizations.
We present the first fully nonlinear study on the accretion of scalar fields onto a seed black hole in anti-de Sitter spacetime in Einstein-Maxwell-scalar theory. Intrinsic critical phenomena in the dynamical transition between the bald and scalarized black holes are disclosed. In scalarizations, the transition is discontinuous and a metastable black hole acts as an attractor at the threshold. We construct a new physical mechanism to dynamically descalarize an isolated scalarized black hole. The first results on  critical phenomena in descalarizations are revealed. The dynamical descalarizations  can be either discontinuous or continuous at the threshold, distinguished by whether or not an intermediate attractor appears.  

%We study the full nonlinear dynamical transition between the  bald black hole and the scalarized black hole in asymptotically anti-de Sitter spacetime. There are two thresholds of the initial scalar data, one for critical scalarisation and one for critical descalarization. For model with nonlinear scalarisation, at both thresholds, the transition is of first order and there is a metastable scalarized black hole playing the role of attractor separating the final bald/scalarized black holes. Interestingly, the dynamical stability does not quantitatively coincide with the thermodynamic stability. For model with spontaneous scalarisation, the initial bald black hole behaves as the critical solution at the scalarisation threshold, though only scalarized black hole survives as the final state. At the descalarization threshold, the transition is of second order and no critical solution appears. We construct a new physical mechanism for the descalarization through black hole accretion. 
\end{abstract}
\maketitle

{\it \textbf{Introduction.}}
The black hole (BH) spontaneous scalarization \citep{Damour1993,Damour1996,Harada1997} has received much attention recently in extended scalar-tensor-Gauss-Bonnet (eSTGB) gravity \citep{Doneva1711,Silva1711,Antoniou1711} and Einstein-Maxwell-scalar (EMS) theory \citep{Herdeiro:2018wub}. Linear tachyonic instability can be triggered in bald BHs and transform them into  scalarized black holes (SBHs). 
%\textbf{It was argued that}
%\red{Yet} 
%a scalarized remnant of spontaneous scalarization 
%in eSTGB gravity 
%\textbf{can be formed in mergers of BH binaries} \cite{Silva:2020omi,2204.}.  
%It was reported that 
Moreover, %it was shown that
the linearly stable bald BHs can also be scalarized, but through a new nonlinear mechanism, different from the spontaneous scalarization \citep{Blazquez-Salcedo:2020nhs,Blazquez-Salcedo:2020crd,Doneva:2021tvn,Liu:2022fxy}. A fully nonlinear dynamical study showed that such scalarization can be realized through the   accretion of scalar field  onto a seed %central  
BH. Novel dynamical critical behaviors at
the threshold of the transition between bald and scalarized BHs were disclosed  \cite{Zhang:2021nnn}, which are reminiscent of the type I critical gravitational collapse \citep{Bizon:1998kq,Choptuik:1996yg}. There exists a metastable SBH that plays the
role of a critical solution separating the final scalarized and bald BHs and behaves as an attractor during the evolution.  

Available studies concentrated on 
%the cases in 
asymptotically flat spacetimes. 
It is known that dynamics in anti-de Sitter (AdS) spacetimes are  different from that in asymptotically flat spacetimes. The turbulent instability in pure AdS spacetimes brings different  structures in critical gravitational collapses from those in flat spaces  \citep{Bizon:2011gg,Choptuik:1992jv,Gundlach:2007gc,Liebling:1996dx,Evans:1994pj}.  Charged and rotating AdS BHs  are  unstable due to superradiant instability, while  their counterparts are stable in asymptotically flat spacetimes \cite{Bosch:2016vcp,Chesler:2018txn,Berti:2009kk,Brito:2015oca}.  These interesting results inspire us to further investigate novel critical phenomena in AdS spacetimes.  

% \cite{Gundlach:2007gc,Liebling:1996dx,Evans:1994pj}? 
%Considering the peculiar spacetime properties, it is interesting to generalize the discussion to anti-de Sitter (AdS) spacetimes. The AdS boundary acts like a mirror that bounces the outgoing energy back into the bulk.  This makes the critical gravitational collapse in  AdS spacetimes \citep{Bizon:2011gg}  qualitatively different from that in asymptotically flat spacetimes \citep{Choptuik:1992jv,Gundlach:2007gc}. 
%In eSTGB theory, this boundary makes the formation of spontaneous scalarization easier in AdS spacetimes  than that in asymptotically flat cases \cite{2012.11844}. 
%In this work we investigate for the first time a complete dynamical formation of SBHs through the nonlinear accretion of   scalar fields onto a seed AdS bald BH and disclose new critical phenomena which are  different from those revealed in asymptotically flat spacetimes \citep{Zhang:2021nnn}. %The final black hole mass of  the supercritical solutions do not follow the power-law scaling anymore.

We explore the dynamical formation of SBHs through  scalar fields accretion onto seed AdS bald BHs and examine the influence of the perturbation strength on  the scalarization. We illustrate special dynamical critical phenomena in the spontaneous transition and find the bald BH playing the role of critical solution.  
%   \citep{Fernandes:2019kmh,Ripley:2020vpk,Kuan:2021lol,East:2021bqk, Zhang:2021etr}.
%We disclose special dynamical critical phenomena in such process that the initial bald BH plays the role of a CS and  only  SBH survives as the  final state.  
After accretion of strong enough scalar perturbations, SBH  can be descalarized.  %by shedding off its   scalar hair and leave a stable bald remnant. 
In addition to  those shown in binary BH mergers   \cite{Silva:2020omi, Doneva:2022byd} and  superradiance \cite{Corelli:2021ikv}, we construct a new physical mechanism for the dynamical descalarization of an  isolated  SBH. We  reveal  the first results on the critical behaviors in dynamical descalarization. % and compare them with those in scalarizations. 
 Unlike the transition at the scalarization threshold which is always discontinuous, the transition at the descalarization threshold  can be either discontinuous or continuous, depending on the specific model.  Furthermore, we reveal more connections between dynamical critical scalarizations/descalarizations and  critical  gravitational collapses in \cite{Gundlach:2007gc,Liebling:1996dx,Evans:1994pj}. 

%critical phenomena in descalarizations  are qualitatively different in different transitions. 

{\it{\textbf{Numerical setup.}}}
The action of the EMS theory in AdS spacetime is 
\begin{equation}
S=\int d^{4}x\sqrt{-g}(R+\frac{6}{L^{2}}-2\nabla_{\mu}\phi\nabla^{\mu}\phi-f(\phi)F_{\mu\nu}F^{\mu\nu}),\label{eq:action}
\end{equation}
where $R$ is the Ricci scalar and   $L$ is the AdS radius. The real scalar field 
$\phi$ couples to the Maxwell field $F_{\mu\nu}\equiv\partial_\mu A_\nu-\partial_\nu A_\mu$  through function $f(\phi)$.
We consider a model with $f=e^{\alpha\phi^{2}}$, which can trigger the linear tachyonic instability in   Reissner-Nordstr\"om (RN)  AdS BHs with large charge to mass
ratio, so that the spontaneous scalarization occurs and the stable solutions become SBHs \citep{Guo:2021zed,Zhang:2021etr}. Further we consider a model with $f=e^{\beta\phi^{4}}$ and show   that the RN-AdS BH is linearly stable against small perturbation, but nonlinearly unstable against large perturbation and evolves into a SBH in AdS spacetime through a new nonlinear mechanism beyond the spontaneous scalarization. Here $\alpha,\beta$ are coupling constants. 
%Note that both models we consider are invariant under  $\phi\to-\phi$. 

To study the nonlinear BH dynamics in spherically symmetric asymptotically AdS spacetime, we take the ingoing
Eddington-Finkelstein coordinate \citep{Chesler:2013lia}:
\begin{equation}
ds^{2}=-Wdt^{2}+2dtdr+\Sigma^{2}(d\theta^{2}+\sin^{2}\theta d\varphi^{2}).
\end{equation} 
Here $W,\Sigma$ are metric functions of $(t,r)$. This coordinate is regular on the BH apparent horizon $r_{h}$ satisfying 
$g^{\mu\nu}\partial_{\mu}\Sigma\partial_{\nu}\Sigma=0%\label{eq:horizon}
$ which implies the vanishing  of    auxiliary variable $S\equiv\partial_{t}\Sigma+\frac{1}{2}W\partial_{r}\Sigma.$
The BH irreducible mass $M_{h}(t)\equiv\sqrt{\frac{V_{h}}{4\pi}}=\Sigma(r_{h},t)$ where $V_h$ is the apparent horizon area. 
It measures the thermodynamic entropy of a BH.  We take the   gauge field 
$A_{\mu}dx^{\mu}=A(t,r)dt$. The Maxwell
equation gives $\partial_{r}A=\frac{Q}{\Sigma^{2}f}.$ Here $Q$
is the BH charge. %The nonvanishing scalar weakens the Maxwell field. 
Introducing auxiliary variable
\begin{equation}
P=\partial_{t}\phi+\frac{1}{2}W\partial_{r}\phi,\label{eq:Pt}
\end{equation}
the Einstein equations give
\begin{align}
\partial_{r}^{2}\Sigma= & -\Sigma(\partial_{r}\phi){}^{2},\label{eq:zr}\\
\partial_{r}S= & \frac{1-2S\partial_{r}\Sigma}{2\Sigma}+\frac{3\Sigma}{2L^{2}}-\frac{Q^{2}}{2\Sigma^{3}f},\label{eq:Sr}\\
\partial_{r}^{2}W= & -4P\partial_{r}\phi+\frac{4S\partial_{r}\Sigma-2}{\Sigma^{2}}+\frac{4Q^{2}}{\Sigma^{4}f},\label{eq:ar}\\
\partial_{t}S= & \frac{S\partial_{r}W-W\partial_{r}S}{2}-\Sigma P^{2}.\label{eq:St}
\end{align}
The scalar equation gives
\begin{equation}
\partial_{r}P=-\frac{P\partial_{r}\Sigma+S\partial_{r}\phi}{\Sigma}-\frac{Q^{2}}{4\Sigma^{4}f{}^{2}}\frac{df}{d\phi}.\label{eq:Pr}
\end{equation}
%Given initial $\phi$, we can successively work out $\Sigma,S,P$ and $W$ from equations (\ref{eq:zr}, \ref{eq:Sr}, \ref{eq:Pr}) and (\ref{eq:ar}). Then from (\ref{eq:Pt}) we get $\phi$ on the next time slice. Iterating this procedure, we get all the variables on all time slices. We specify reflecting boundary conditions at the AdS boundary. 
The asymptotic behaviors of the variables are
\begin{align}
\phi=&\frac{\phi_{3}(t)}{r^{3}}+O(r^{-4}),\label{eq:phi}
\\\Sigma=&r+\lambda(t)-\frac{3\phi_{3}(t)^{2}}{10r^{5}}+O(r^{-6}),\label{eq:Sigma}
\\S=&\frac{(r+\lambda(t))^{2}}{2L^{2}}+\frac{1}{2}-\frac{M}{r}+O(r^{-2}),\label{eq:S}
\\P=&-\frac{3\phi_{3}(t)}{2L^{2}r^{2}}+O(r^{-3}),\label{eq:P}
\\W=&\frac{(r+\lambda(t))^{2}}{L^{2}}-2\dot{\lambda}(t)+1-\frac{2M}{r}+O(r^{-2}),\label{eq:W}
\end{align}
Here the constant $M$ represents the  ADM mass. %, which contains all the energy of the metric, the Maxwell field and the scalar field.
$\phi_{3}(t)$ is an unknown function of time that is determined
by the dynamics. $\lambda(t)$ is a residual gauge freedom  \cite{Chesler:2018txn,Chesler:2013lia} and the
dot denotes the time derivative. 
%Using this gauge and (\ref{eq:St}), we can fix the position of the apparent horizon at $r_{h}=1$   during the evolution \citep{Chesler:2013lia}. 
The procedure to solve these equations, the boundary conditions and the test of the convergence are shown in the supplement material. %The numerical accuracy and convergence are checked by using different grid points. 

Hereafter we fix $L=1$ such that all physical quantities are measured by the AdS radius.  Without loss of generality,  we choose the initial
configuration as a seed  RN-AdS BH with   $M_{0}=1,Q=0.8$ and simulate the evolution under initial scalar data $\phi_{0}=pe^{-64(0.5-\frac{1}{r})^{2}}$.
The dynamical critical behaviors for both scalarization and descalarization are not changed qualitatively for different parameters and  initial scalar data families. The ADM mass $M$ increases
with   $p$ such that the evolution of the system can be viewed as the   accretion of  scalar field  onto the     central BH.  
%We also simulate the evolution with other initial scalar data families.
 % such as $\phi_{0}=p(\frac{1}{r}-0.4)^{2}(\frac{1}{r}-0.8)^{2}$ if $\frac{1}{r}\in[0.4,0.8]$ and zero otherwise. 
%The results are not changed qualitatively. 
%We adopt the numerical approach described in  \citep{Chesler:2013lia}. The numerical accuracy and convergence are checked by using different grid points. 

For any initial data and coupling functions, we find universal and robust relations
during the  evolution, 
\begin{equation}
\ln\dot{M}_{h}=2\ln|\dot{\phi}_{h}|+\text{const}.\label{eq:twice1}
\end{equation}
Here $\phi_{h}(t)$ is the scalar value on the apparent horizon. $M_{h}$ never decreases in evolution due to the second law of black hole mechanics.
These relations can be partially understood from the perturbative
analysis for stationary solutions, in which $S,W,P|_{r_{h}}=0$ on the horizon.
Combining (\ref{eq:Sr},\ref{eq:St},\ref{eq:Pr}), there are $\partial_{t}S,\partial_{t}\Sigma|_{r_{h}}\propto\delta\phi^{2}$.
Since $M_{h}=\Sigma(r_{h},t)$ and $r_{h}$ satisfies $S(r_{h},t)=0$,
we have $\dot{M}_{h}=-\frac{\partial_{t}S}{\partial_{r}S}\partial_{r}\Sigma|_{r_{h}}+\partial_{t}\Sigma|_{r_{h}}\propto\dot{\phi}_{h}^{2}$.
Thus (\ref{eq:twice1}) is expected for a solution which
can be approximated by a stationary solution.
However, (\ref{eq:twice1}) holds even in the highly nonlinear
regime and calls for further exploration \cite{Zhang:2021nnn,Zhang:2021etr,Zhang:2021edm,Zhang:2021ybj,Luo:2022roz}.

\begin{figure}[h]
\begin{centering}
\includegraphics[width=0.8\linewidth]{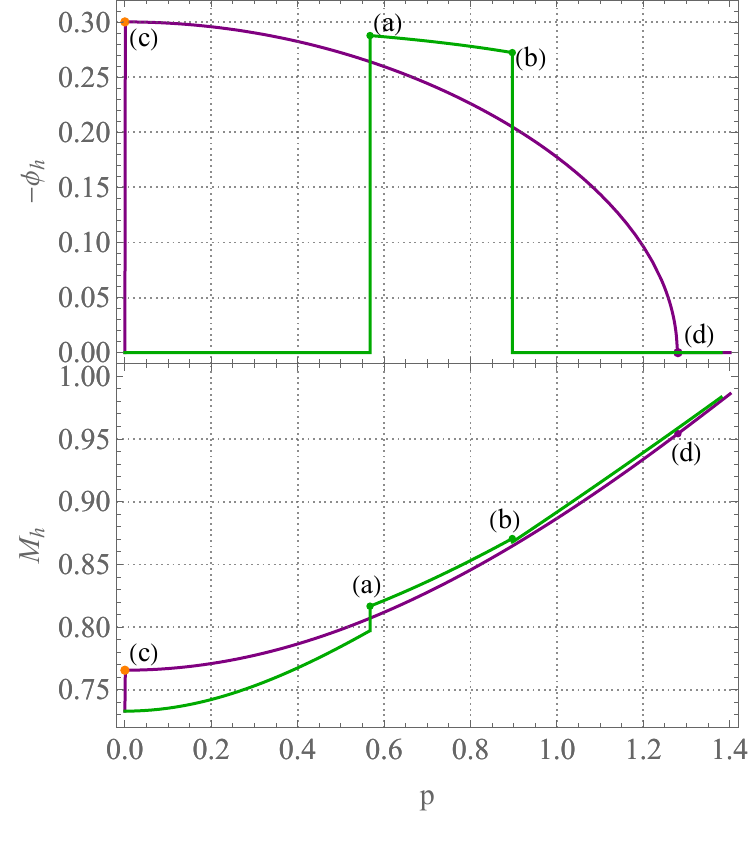}
\par\end{centering}
{\footnotesize{}\caption{{\footnotesize{}\label{fig:Mp}The final values of $\phi_{h}$  and $M_{h}$ versus $p$ of initial data $\phi_{0}=pe^{-64(0.5-\frac{1}{r})^{2}}$.
The green and purple lines are results for $f(\phi)=e^{500\phi^{4}},e^{15\phi^{2}}$
respectively. {The system begins to undergo the process of scalarization and descalarization at the critical points $(p_{a}=0.56736843744772\pm10^{-14},p_c=0)$ and $(p_{b}=0.89673160782993\pm10^{-14},p_{d}=1.2784440\pm10^{-7})$, respectively. Note that the system needs very long time to reach equilibrium near $p_{d}$, so it is more expensive to calculate until the high accuracy $10^{-14}$. Moreover, the critical behavior has shown up when we reach the accuracy $10^{-7}$ and will not change if we further improve the accuracy.} }}
}{\footnotesize\par}
\end{figure}

\begin{figure}[h]
\begin{centering}
\includegraphics[width=0.8\linewidth]{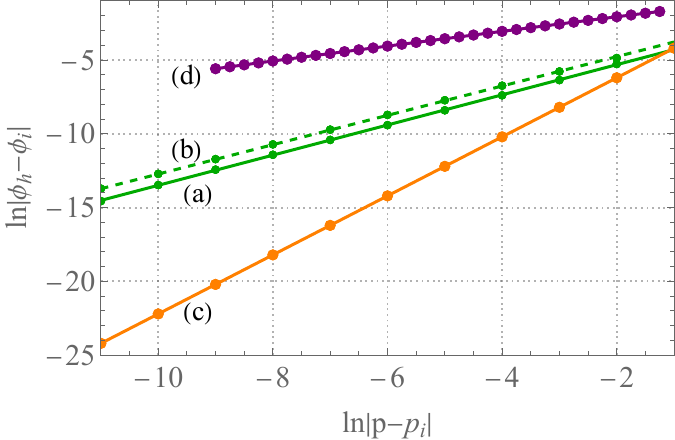}
\par\end{centering}
{\footnotesize{}\caption{{\footnotesize{}\label{fig:philnp}The final $\phi_h$ of SBHs versus 
$p$ near the thresholds. The solid/dashed green, orange and purple
lines are for those near  $p_{a,b,c,d}$ in Fig.\ref{fig:Mp}, respectively.}}
}{\footnotesize\par}
\end{figure}

{\it{\textbf{Numerical results for scalarization.}}} 
We show the final values of  $M_{h}$ and $\phi_{h}$ with respect
to parameter $p$ when the system reaches equilibrium after the initial perturbation in Fig.\ref{fig:Mp}.  For a model with $f=e^{15\phi^{2}}$, both $\phi_{h}$ and $M_{h}$  jump %at arbitrarily small threshold value
at threshold 
$p_c=0$.
The nonzero final $\phi_h$ marks the formation of a  SBH.  The initial RN-AdS BH is spontaneously scalarized due to the   tachyonic instability.  
For a model with $f=e^{500\phi^{4}}$, the initial RN-AdS BH is linearly stable and all small perturbations are absorbed, resulting   a larger scalar-free RN-AdS BH.  But it becomes nonlinearly unstable when $p$ is larger than  the threshold $p_a$, where both $\phi_{h}$ and $M_{h}$ jump. The nonlinear instability indicates a new mechanism for the AdS BH scalarization that is different from the spontaneous scalarization  \cite{Zhang:2021etr}. %To destroy the RN-AdS BH, one needs big enough perturbation. 
This is reflected in the different scaling   of $\phi_h$ for SBHs near thresholds $p_{a,c}$, as shown in Fig.\ref{fig:philnp}, 
\begin{equation}
|\phi_h-\phi_{i}|\propto|p-p_{i}|^{\gamma_{i}}.\label{eq:gamma}
\end{equation}
Here subscript $i\in {a,c}$ and $\phi_{i}$
is the final $\phi_{h}$ for threshold $p_i$. The exponent $\gamma_{a,c}=1,2$ are independent of the initial scalar data family,
the   BH parameters and the coupling parameters $\alpha,\beta$. 
%in the coupling function. 
{However, the nontrivial power law for the BH irreducible mass like in asymptotically flat spacetime \cite{Zhang:2021nnn} is absence here, due to the confining AdS boundary which prohibits energy escape.} 

\begin{figure}
\begin{centering}
\begin{tabular}{cc}
\includegraphics[width=0.5\linewidth]{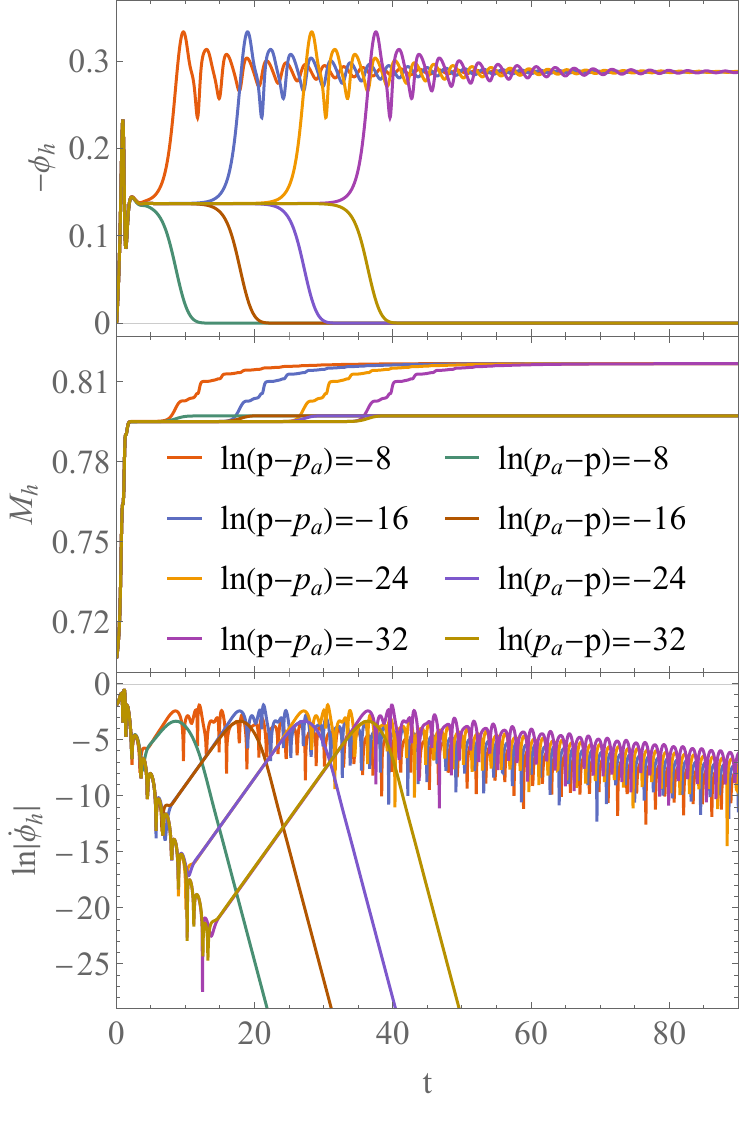} & \includegraphics[width=0.5\linewidth]{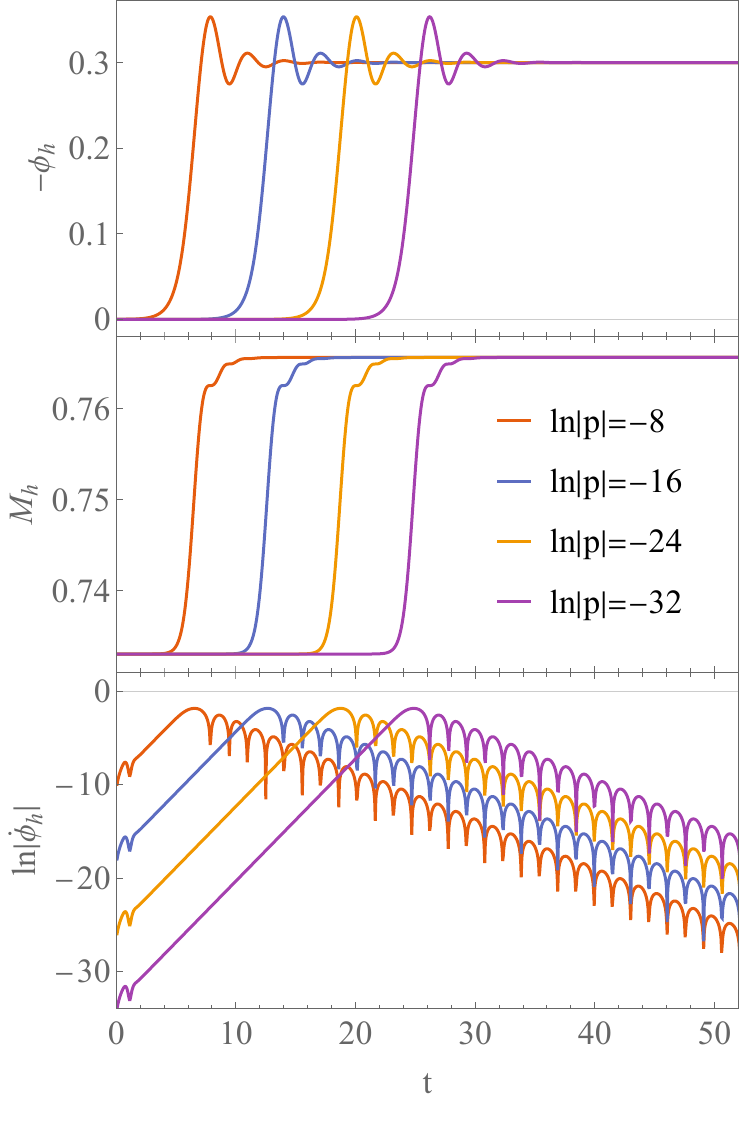}\tabularnewline
(a) & (c)\tabularnewline
\end{tabular}
\par\end{centering}
{\footnotesize{}\caption{{\footnotesize{}\label{fig:MphiptAC}The evolution of the $M_{h},\phi_{h}$,
and $\ln|\dot{\phi}_{h}|$ near the thresholds $p_{a}$ (left) and
$p_{c}$ (right). }}
}{\footnotesize\par}
\end{figure}

\begin{figure}[h]
\begin{centering}
\includegraphics[width=0.8\linewidth]{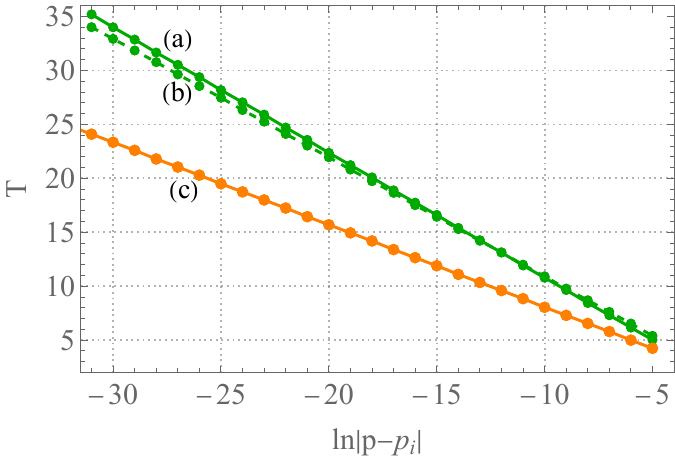}
\par\end{centering}
{\footnotesize{}\caption{{\footnotesize{}\label{fig:Tln}The time $T$ of the intermediate
solutions staying near the CSs versus $\ln|p-p_{i}|$. The
solid, dashed green and orange lines are for thresholds $p_{a,b,c}$, respectively.}}
}{\footnotesize\par}
\end{figure}

In Fig.\ref{fig:MphiptAC}(a) we display the   evolution of $\phi_h,M_h$ and $\ln |\dot{\phi}_h|$ for the model with $f=e^{500\phi^{4}}$ near threshold $p_a$. All  solutions are attracted to an intermediate plateau corresponding to a critical solution (CS), which is a metastable SBH behaving as an attractor. At late times, the solutions decay to bald AdS BHs if $p<p_a$, or to SBHs if $p>p_a$. The evolution can be divided into three stages, as refined in the bottom of Fig.\ref{fig:MphiptAC}(a). At the first stage, the solutions converge to the CS with the   dominant mode frequency $\omega_{a}^{c}=2.81-1.61i$, where the superscript $c$ denotes the CS. At the second stage, the solutions depart   the CS with $|\dot{\phi}_{h}|\propto e^{\eta_{a}t}$ in which $\eta_{a}=0.861$.  The time  $T_{a}$ of the intermediate solutions (the first and second stages) stay near the plateau scales as
\begin{equation}
T_{a}=-\eta_{a}^{-1}\ln|p-p_{a}|+\text{const},\label{eq:Tlnp}
\end{equation}
as shown in Fig.\ref{fig:Tln}. Based on these observations, we conclude that the intermediate solutions $\phi_{p}(t,r)$  near the threshold $p_{a}$  can be approximated by 
\begin{equation}
\phi_{p}(t,r)\approx\phi_{a}(r)+(p-p_{a})e^{\eta_{a}t}\delta\phi_{a}(r)+\text{stable modes}.\label{eq:c1}
\end{equation}
Here $\delta\phi_{a}(r)$ is the  unstable eigenmode with
eigenvalue $\eta_{a}$ for the CS $\phi_{a}(r)$. The unstable mode
grows to a finite size in time $T_{a}$, after which the solutions converge either to  larger bald RN-AdS BHs 
with dominant mode $\omega_{a}^{b}=-2.33i$ if $p<p_a$, or to
SBHs with dominant mode $\omega_{a}^{s}=2.31-0.0649i$  if $p>p_a$. Superscripts $b,s$ denote bald, scalarized BHs respectively.

In Fig.\ref{fig:MphiptAC}(c) we show the   evolution  of $\phi_h,M_h$ and $\ln |\dot{\phi}_h|$ for the model with $f=e^{15\phi^{2}}$ near   threshold $p_c$. They look qualitatively similar to those in Fig.\ref{fig:MphiptAC}(a), except some intriguing differences.  Here only SBH survives as the final state and the initial RN-AdS BH plays the role of a CS.  %This is consistent with the fact the RN-AdS BH  suffers the linear  tachyonic instability. % which happens just outside the  horizon  and is not altered by the AdS boundary. 
%\textbf{We exhibit that any arbitrarily small perturbation near the horizon of the original bald BH seed can trigger the transition to transform the bald BH to the scalarized one. The AdS boundary cannot influence such transformation. }
The intermediate solution stays near   it with   time   $T_{c}\propto-\eta_{c}^{-1}\ln|p|$
where $\eta_{c}=1.31$, as shown in Fig.\ref{fig:Tln}. The evolution here is actually a special case of (\ref{eq:c1}) with threshold $p_c=0$ and CS $\phi_c(r)=0$.
%$\eta_{c}$ depends on the initial spacetime parameters and the coupling parameter $\beta$, but is independent of the initial scalar data families. 
After $T_c$, the solution converges to the final SBH with  dominant mode  $\omega_{c}^{s}=2.06-0.524i$. % at late times.

{\it{\textbf{Numerical results for descalarization.}}}
%In eSTGB theory, the dynamical descalarization was discovered in binary BH mergers. 
%In eSTGB theory, it was discovered that after the merger of a SBH with the other bald or scalarized BH, the scalar field starts dissipating and the final BH could be descalarized \cite{Silva:2020omi, Doneva:2022byd}. Such a process was expected to  leave imprint in gravitational wave observations. 
The dynamical descalarization induced by  binary BH mergers in eSTGB theory  \cite{Silva:2020omi, Doneva:2022byd} and superradiance in EMS theory \cite{Corelli:2021ikv} has been reported. 
%Here we present the first study on the dynamical descalarization for an isolated  SBH  in the EMS theory and 
Here we show that when the energy of the initial perturbation is large enough,  
the accretion of  scalar field onto an isolated SBH can also trigger the scalar field dissipation and descalarize the SBH. 
As shown in Fig.\ref{fig:Mp}, we observe that $\phi_h$ gradually or suddenly drops to zero at thresholds $p_{d,b}$ for models with $f=e^{15\phi^{2}},e^{500\phi^{4}}$, respectively. These indicate two   kinds of phenomena in the dynamical descalarization through    accretion, which are continuous or discontinuous at thresholds $p_{d,b}$, respectively. 
 The scaling law  (\ref{eq:gamma}) still hold here, but with universal exponents $\gamma_{d,b}=1,0.5$  respectively. 
{The reason for this significant difference in the dynamic processes is the different phase structures possessed by the two models. This will be described from the perspective of static solutions in the next section.}

\begin{figure}
\begin{centering}
\begin{tabular}{cc}
\includegraphics[width=0.5\linewidth]{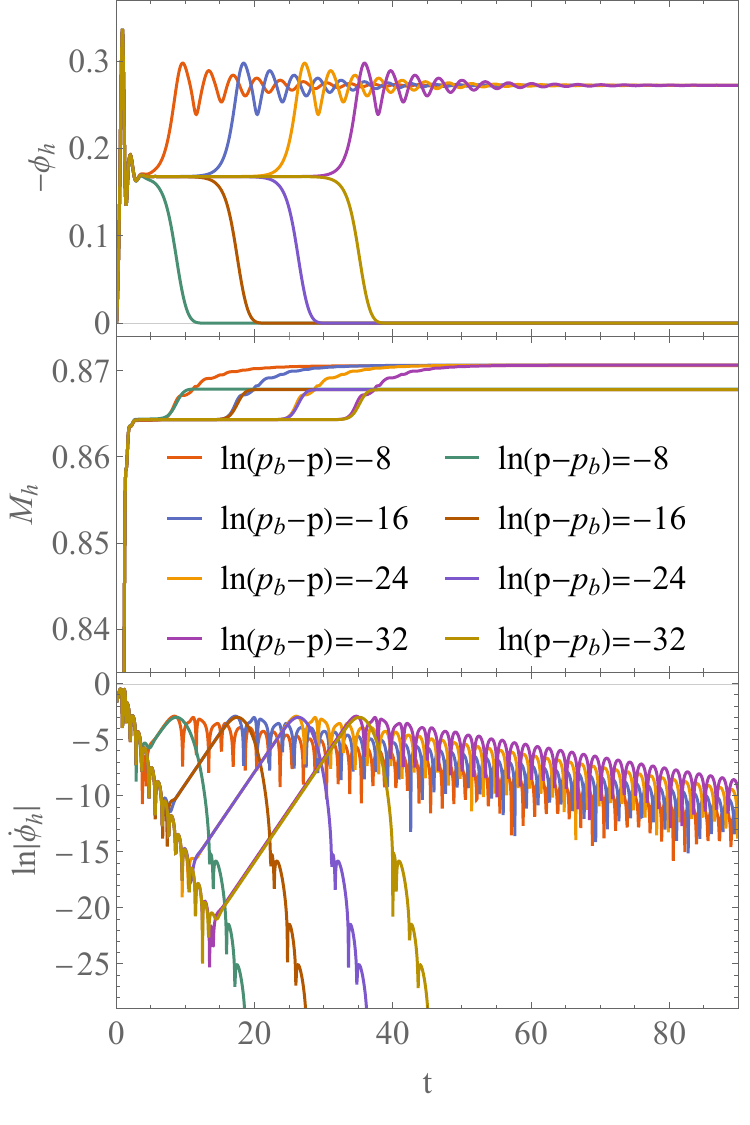} & \includegraphics[width=0.5\linewidth]{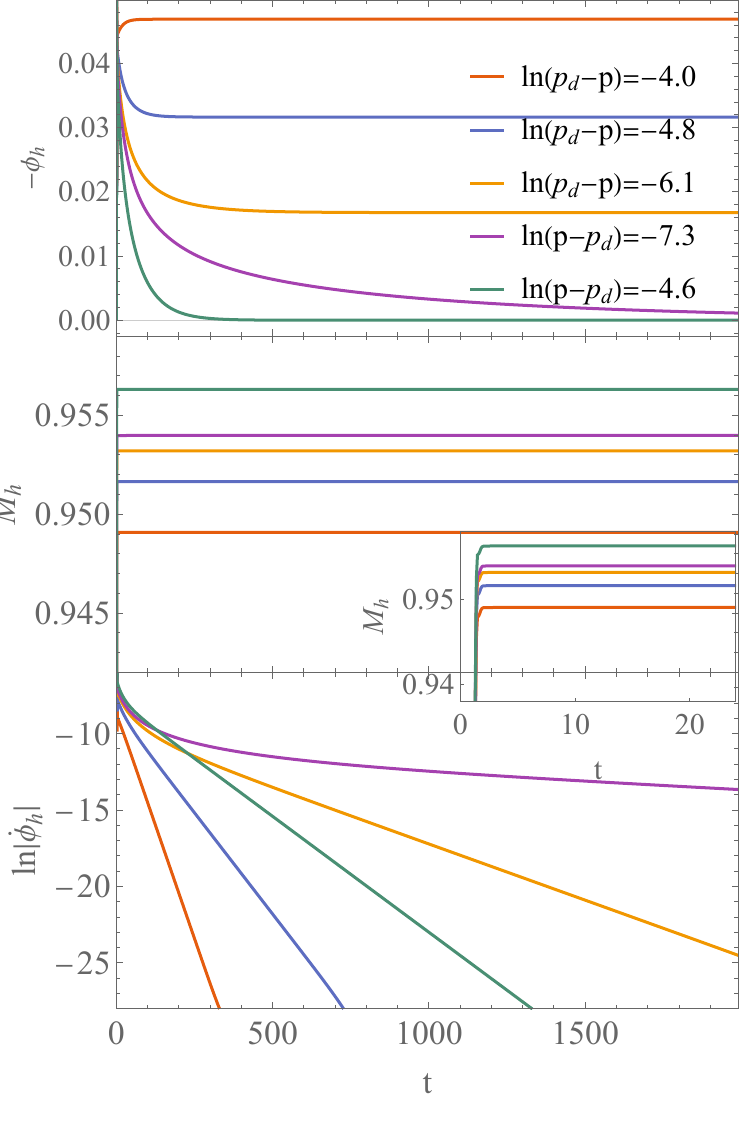}\tabularnewline
(b) & (d)\tabularnewline
\end{tabular}
\par\end{centering}
{\footnotesize{}\caption{{\footnotesize{}\label{fig:MphiptBD}The evolution of the $M_{h},\phi_{h}$,
and $\ln|\dot{\phi}_{h}|$ near the thresholds $p_{b}$ (left) and
$p_{d}$ (right).}}
}{\footnotesize\par}
\end{figure}

%Through nonlinear numerical simulation, we disclose critical phenomena in descalarizations.
As shown in  Fig.\ref{fig:MphiptBD}(b), the evolution near the descalarization threshold $p_b$ is qualitatively the same with that near the scalarization threshold $p_a$. The intermediate solutions are attracted to a CS with dominant mode $\omega_{b}^{c}=3.25-1.55i$. Then the solutions depart   the CS with $|\dot{\phi}_{h}|\propto e^{\eta_{b}t}$ where $\eta_{b}=0.906$. The  time $T_b$ of the intermediate solution that stays near the CS scales as $T_{b}\propto-\eta_{b}^{-1}\ln|p-p_{b}|$, as exhibited in Fig.\ref{fig:Tln}. After $T_b$, the solution keeps scalarized with dominant mode $\omega_{b}^{s}=1.88-0.104i$ if $p<p_b$, or decays to a bald BH with dominant mode  $\omega_{b}^{b}=2.21-3.01i$ if $p>p_b$.

The critical phenomena of the descalarization near  threshold $p_d$ are special, as shown in   Fig.\ref{fig:MphiptBD}(d). There is no attractor at the threshold. Given $p$, we find that the scalar field    exponentially decays at late times with the same rate  in the whole space. 
%This dynamical  feature together with the universal power law (\ref{eq:gamma}) with $\gamma_d=0.5$   resembles the type II critical gravitational collapse with continuous self-similarity \cite{Gundlach:2007gc,Evans:1994pj,Liebling:1996dx}.
However, for different $p$, the late time exponential decay rate varies with $p$ as  $\omega_{d}^{s}=(3.20p-4.09)i$  for the final SBH  when $p<p_d$,  or  $\omega_{d}^{b}=(2.01-1.57p)i$   for the final bald RN-AdS BH when $p>p_d$.   Both  $\omega_{d}^{s}$ and $\omega_{d}^{b}$ are zero at threshold $p_d$.   This is also qualitatively different from those near $p_{a,b,c}$ which are independent of $p$. 

 %After $p_d$, we always get the stable bald remnants as shown in Fig.\ref{fig:Mp}. 

\begin{figure}[h]
\begin{centering}
\includegraphics[width=0.9\linewidth]{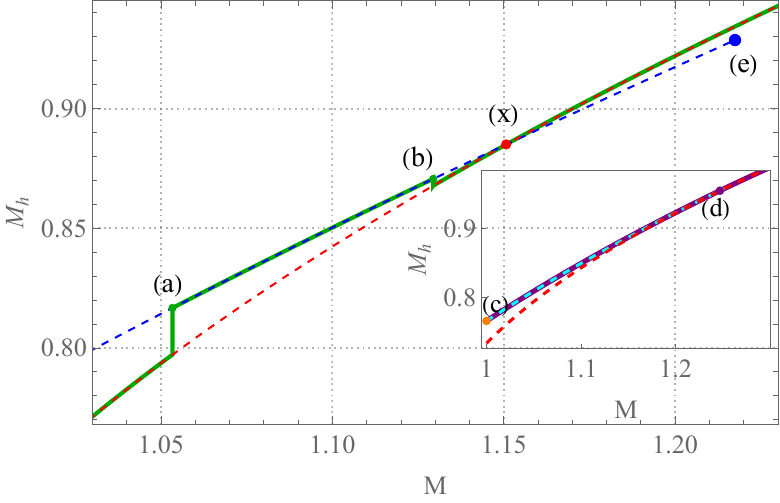}
\par\end{centering}
{\footnotesize{}\caption{{\footnotesize{}\label{fig:SM}The black hole irreducible mass versus
the ADM mass for static solutions. The dashed red lines are for
the static RN-AdS solutions. The dashed blue and cyan (inset) lines for the static SBH solutions in models with $f=e^{500\phi^{4}},e^{15\phi^{2}}$,
respectively. The intersection points (x)$=$(1.1507, 0.8851) and
(d)$=$(1.2469, 0.9542). There are no static SBHs beyond  (e)=(1.2175,
0.9287) for $f=e^{500\phi^{4}}$, or  (d) for $f=e^{15\phi^{2}}$.
The solid green and purple lines correspond to those in Fig.\ref{fig:Mp} obtained
in nonlinear evolution.}}
}{\footnotesize\par}
\end{figure}

{\it{\textbf{Static solutions and thermodynamics.}}}
{In Fig.\ref{fig:SM} we illustrate  the BH irreducible mass versus the ADM mass for the two branches of  static  solutions: the RN-AdS BHs and SBHs. For the model with $f=e^{15\phi^2}$ shown in the inset, the static SBH smoothly intersects the static RN-AdS BH at point (d). Before point (d), the RN-AdS BH is unstable under arbitrarily small perturbation and dynamically evolves to the SBH which is also favored by thermodynamics. After point (d), only the static RN-AdS BH solutions survives and it becomes  stable under small perturbation.
The final state of the dynamical evolution turns from the SBH to the RN-AdS BH at point (d). Combining with Fig.\ref{fig:Mp} we see that here the descalarization is a continuous phase transition.  %Therefore, there is only one path from stable SBH across point (d) to stable RN-AdS BH through accretion process, where the continuous transition will occur.
For  the model with $f=e^{500\phi^4}$, the RN-AdS and the SBHs  coexist before   point (e). But now  they both are linearly stable. So although the SBH has larger or smaller entropy than the RN-AdS BH before or after point (x), a small  perturbation  does not necessarily lead to a dynamical transition between the two branches of stable solutions. However, with a sufficiently large 
quench, the system can be interconverted between these two stable states. The dynamic process is a first-order phase transition and does not generically coincide with the equilibrium thermodynamic preference for static solutions. } 

{\it{\textbf{Summary and discussion.}}}
  %we disclosed that the scalarization can be triggered at any arbitrarily small perturbation, 
We have studied for the first time the scalarization in AdS spacetime  through nonlinear accretion of scalar field onto a central BH. In the dynamical spontaneous scalarization, the RN-AdS BH plays the role of a CS and only SBH survives as the final state. For a new nonlinear mechanism beyond the spontaneous scalarization, we revealed that  a linearly stable bald BH can be transformed into a SBH, provided that the perturbation strength is over a threshold.  Near the threshold, a metastable SBH in AdS spacetime acts as an attractor separating   the final bald and scalarized BHs. 
The transitions at the scalarization thresholds in both cases are of the first order   resembling the type I critical collapse  \cite{Bizon:1998kq,Choptuik:1996yg}. However, due to the confining AdS boundary, the final BH mass does not follow a nontrivial power-law  as found in asymptotically flat spacetime  \cite{Zhang:2021nnn}.

We  have constructed a new physical mechanism for the dynamical descalarization of an isolated SBH through   accretion of scalar field.  
For a model with $f=e^{\alpha\phi^2}$, the final BH mass and scalar field  changes continuously with  a nontrivial universal power-law at the descalarization threshold, 
%but no attractor appears. The scalar field exhibits local self-similarity and   
resembling the type II critical collapse \citep{Gundlach:2007gc,Evans:1994pj,Liebling:1996dx}. No
attractor appears during the dynamical descalarization. 
For $f=e^{\beta\phi^4}$, we observe  sudden drops in both the final scalar field and BH irreducible mass  at the descalarization threshold, indicating a first order transition. There is an attractor separating the final scalarized and bald BHs in such transition resembling the type I critical collapse.

% final BH mass changes smoothly, indicating a continuous transition in which  at .  The 

%If we take $f=e^{10\phi^2}$, there is no attractor at the descalarization threshold, but the final scalar value on the BH horizon obeys a power-law.  The scalar field gradually dissipating from the hairy hole, and there appears a smooth change of  $M_h$ at the critical point $p_d$.  The AdS SBH is transformed into a larger bald AdS hole so that $M_h$ grows and the second law of thermodynamics keeps. However,   taking $f=e^{500\phi^4}$, we observe a sudden drop of   BH mass at the threshold $p_b$. Including the energy drained by the descalarization, the second law of thermodynamics is still respected at $p_b$. Although $M_h$ drops, the ADM mass keeps growing. 

The dynamical critical behaviors we disclosed in this work are not limited in the EMS theory. The first-order strong gravity phase  transitions exist  in other physical models, such as the eSTGB theory \cite{Doneva:2021tvn,Liu:2022fxy}, the BH with Q-cloud \cite{ Herdeiro:2020xmb,Hong:2020miv}, black ring \cite{ Emparan:2001wn,Emparan:2007wm},    holographic models such as the holographic QCD  etc \cite{Gubser:2008ny,Janik:2017ykj,Attems:2019yqn}.
They also resemble the first-order matter phase transition in neutron star binary mergers  \cite{Most:2018eaw,Zha:2020gjw}. %The abrupt change in the first-order phase transition is important   in the study of gravitational waves. 
It is expected that the study of dynamical critical behaviors in our work can shed lights
into deep investigations of dynamical mechanisms in many
alternative gravity models. Uncovering these mechanisms
can provide better understanding on the gravitational wave
emission. 

{\it{\textbf{Acknowledgments.} }}
This research is supported by National Key R\&D Program of China under
Grant No.2020YFC2201400, and the Natural Science Foundation of China
under Grant Nos. 11975235, 12005077, 12035016 and Guangdong Basic and Applied Basic Research
Foundation under Grant No. 2021A1515012374. B. W. was partially supported by NNSFC under grant 12075202.

\section*{supplement material}

Besides those presented in the main text, we actually did many calculations with different initial values of $M_{0},Q$ and the parameters $\alpha,\beta$ in coupling functions $f=e^{\alpha\phi^{2}}$ or $e^{\beta\phi^{4}}$.  We found that only for large $\alpha,\beta$ and $Q/M_{0}$, the initial RN-AdS black hole can be scalarized. But the dynamical critical behaviors for both scalarization and descalarization are not changed qualitatively for different parameters and  initial scalar data families. Thus our results reported in the main text are sound. 

Now we explain in details about our numerical method, boundary conditions and our test of convergence. 
We follow an efficient integration strategy  \cite{Chesler:2013lia} suitable  for our works.
%for the set of equations (\ref{eq:Pt}-\ref{eq:Pr})
%(\ref{eq:Pt}, \ref{eq:zr}, \ref{eq:Sr}, \ref{eq:ar}, \ref{eq:St}, \ref{eq:Pr})
% with a nested structure  \cite{Chesler:2013lia}. %We describe it for our works in the following. 

 Given $\phi$,  equation (\ref{eq:zr}) is a linear second order radial ordinary
differential equation (ODE) for $\Sigma$. The two   integration constants
are fixed by the first and second terms in the asymptotic behavior
(\ref{eq:Sigma}): $\Sigma\sim r+\lambda$. Here the initial gauge
$\lambda$ is determined by the radial position of the initial apparent
horizon. For example, if the initial configuration is a RN-AdS BH with $S(r')=\frac{r'^{2}}{2L^{2}}+\frac{1}{2}-\frac{M}{r'}+\frac{Q^{2}}{2r'^{2}}$,
then the initial $\lambda$ is determined by
shifting the radius of horizon $r'_{h}$ to the boundary of computing domain $r_{h}=r'_{h}-\lambda=1$.

Once $\phi$ and $\Sigma$ are known, equation (\ref{eq:Sr})
can be integrated  to solve 
$S$. In our   simulation, the apparent horizon condition
$S(r_{h})=0$ is used to fix the single integration constant in (\ref{eq:Sr})  (Another option is to use the ADM mass $M$ as the integration
constant. We do not adopt this option here).

%Once $\phi$ and $\Sigma$ are known, the equation (\ref{eq:Sr}) can be integrated with the appropriate boundary condition to determine $S$. In our numerical simulation, the apparent horizon condition $S(r_{h})=0$ is used to fix the single integration constant in (\ref{eq:Sr}) for $S$  (Another option is to use the ADM mass $M$ as the integration constant. We do not adopt this option here).

 For equation (\ref{eq:Pr}), the single needed integration
constant can be fixed by the coefficient of the leading term in (\ref{eq:P}):
$P\sim-\frac{3\phi_{3}}{2L^{2}r^{2}}$, where  $\phi_{3}$
can be extracted from the asymptotic behavior of the already known
scalar field (\ref{eq:phi}). 

Equation (\ref{eq:ar}) is a linear second order   ODE for $W$,
whose source term depends only on the solved variables $(\phi,\Sigma,S,P)$. We choose the gauge so that the position of the apparent horizon is time invariant, which  implies $\partial_t S|_{r_h}=0$ and gives a boundary condition for $W$ at the   horizon from   (\ref{eq:St}): 
\begin{equation}
W(r_{h})=-\left.\frac{2\Sigma P^{2}}{\partial_{r}S}\right|_{r_{h}}.\label{eq:Wh}
\end{equation}
 Besides the boundary condition (\ref{eq:Wh})
fixing one of the two integration constants, another
integration constant is fixed by the leading term in the asymptotic
behavior (\ref{eq:W}), $W\sim r^{2}/L^{2}$.

Once $W$ is determined, one can extract the
time derivatives of scalar field $\phi$ and shift $\lambda$ from
the auxiliary variable $P$ (\ref{eq:Pt}) and the asymptotic behavior of $W$ (\ref{eq:W}),
respectively.
\begin{align}
\partial_{t}\phi & =P-\frac{1}{2}W\partial_{r}\phi,\label{eq:phit}\\
\partial_{t}\lambda & =-\frac{1}{2}\lim\limits _{r\rightarrow\infty}\left[W-\frac{(r+\lambda)^{2}}{L^{2}}-1\right].\label{eq:lambdat}
\end{align}
Then one can push the scalar field $\phi$ and shift $\lambda$ to
the next time slice by integrating in time for  (\ref{eq:phit},
\ref{eq:lambdat}).
The procedure is iterated until the entire simulation is completed. 

To improve the stability and accuracy, we introduce new variables $g,\sigma,s,p,a$  in the numerical code 
as follows.
\begin{align}
\phi\equiv & \frac{g}{r^{3}}, 
\Sigma\equiv  r+\lambda+\frac{1}{r^{3}}\sigma,\\
S\equiv & \frac{(r+\lambda)^{2}}{2L^{2}}+\frac{1}{2}+\frac{1}{r}s,\\
P\equiv & -\frac{3\phi_{3}}{2L^{2}r^{2}}+\frac{1}{r^{3}}p,\\
W\equiv & \frac{(r+\lambda)^{2}}{L^{2}}+1+a.
\end{align}
 The boundary conditions should be reorganized to apply
to $g,\sigma,s,p,a$. 
We compactify the radial coordinate by $z=1/r$ such that
$z=1$ corresponds to the apparent horizon and $z=0$ to the AdS boundary.
The new radial coordinate $z$ is discretized with $N$ Chebyshev--Gauss--Lobatto
points $z_{m}=\frac{1}{2}(1+\cos\frac{n\pi}{N})$ for $n=0,1,...,N$. The constraint
equations (\ref{eq:zr}-\ref{eq:Pr})
%(\ref{eq:zr}, \ref{eq:Sr}, \ref{eq:ar}, \ref{eq:Pr})
are solved with collocation approach. The evolution equations (\ref{eq:phit},
\ref{eq:lambdat}) are solved with classic fourth order Runge-Kutta
method.

\begin{figure}[h]
\begin{centering}
\begin{tabular}{cc}
\includegraphics[width=0.5\linewidth]{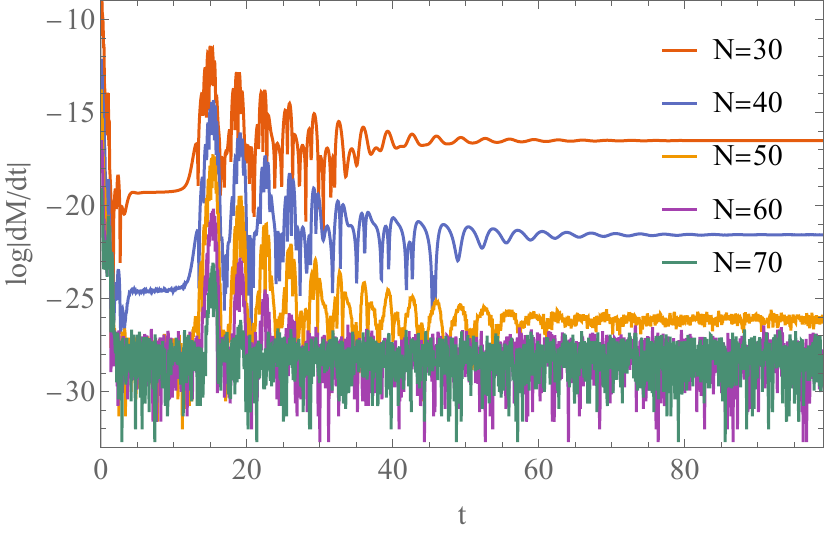} & \includegraphics[width=0.5\linewidth]{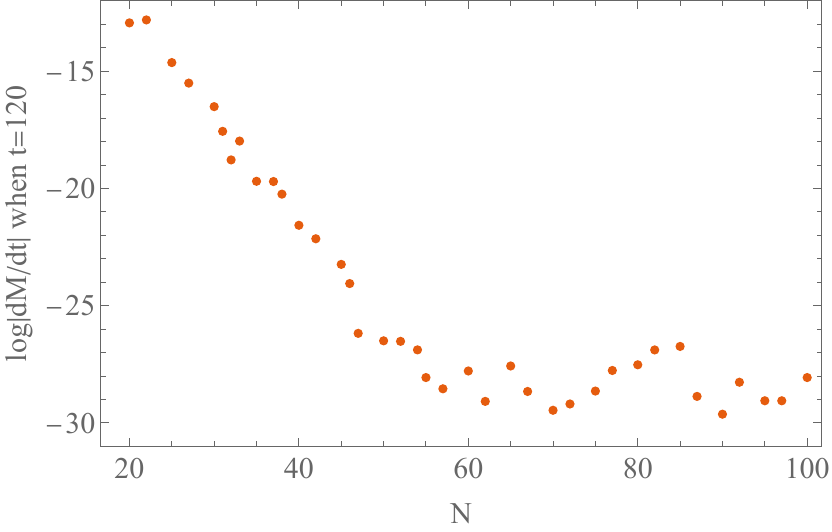}\tabularnewline
\end{tabular}
\par\end{centering}
{\footnotesize{}\caption{{\footnotesize{}\label{fig:err}The accuracy and convergence of our
numerical code. Left: the evolution of $\log|\partial_{t}M|$ with
different grid points $N$. Right: the value of $\log|\partial_{t}M|$
when $t=120$ for different $N$. Here we show the results of the
evolution starting with a RN-AdS balck hole with $M_{0}=1,Q=0.8$
in the EMS model with $f=e^{500\phi^{4}}$ under perturbation $\phi_{0}=pe^{-64(0.5-\frac{1}{r})^{2}}$
with $p=0.89673$.}}
}{\footnotesize\par}
\end{figure}
The redundant equation (\ref{eq:St})   allows us to detect
the numerical errors during the evolution (it reduces to (\ref{eq:Wh})
at the apparent horizon and gives one  boundary condition 
for $W$). Without loss of generality, we use the asymptotic behavior
of equation (\ref{eq:St}) at the AdS boundary, which generates the
energy conservation condition $\partial_{t}M=0$, to show the numerical
error. Fig.\ref{fig:err} shows that the accuracy of our code
improves exponentially as the grid points increase when $N\lesssim60$.
All results shown in the paper are obtained with $N=100$.


\begin{thebibliography}{10}
\bibitem{Damour1993}T. Damour and G. Esposito-Farese, Nonperturbative
strong field effects in tensor-scalar theories of gravitation, Phys.
Rev. Lett., vol. 70, pp. 2220-2223,

\bibitem{Damour1996}T. Damour and G. Esposito-Farese, \textquotedblleft Tensor-scalar
gravity and binary pulsar experiments,\textquotedblright{} Phys. Rev.
D 54 (1996), 1474-1491 {[}arXiv:gr-qc/9602056 {[}gr-qc{]}{]}.

\bibitem{Harada1997}T. Harada, \textquotedblleft Stability analysis
of spherically symmetric star in scalar-tensor theories of gravity,\textquotedblright{}
Prog. Theor. Phys. 98 (1997), 359-379 {[}arXiv:gr-qc/9706014 {[}gr-qc{]}{]}. 

\bibitem{Doneva1711}D. D. Doneva and S. S. Yazadjiev, New Gauss-Bonnet
Black Holes with Curvature-Induced Scalarization in Extended Scalar-Tensor
Theories, Phys. Rev. Lett. 120, no.13, 131103 (2018) {[}arXiv:1711.01187
{[}gr-qc{]}{]}. 

\bibitem{Silva1711}H. O. Silva, J. Sakstein, L. Gualtieri, T. P.
Sotiriou and E. Berti, Spontaneous scalarization of black holes and
compact stars from a Gauss-Bonnet coupling, Phys. Rev. Lett. 120,
no.13, 131104 (2018) {[}arXiv:1711.02080 {[}gr-qc{]}{]}.

\bibitem{Antoniou1711}G. Antoniou, A. Bakopoulos and P. Kanti, Evasion
of No-Hair Theorems and Novel Black-Hole Solutions in Gauss-Bonnet
Theories, Phys. Rev. Lett. 120, no.13, 131102 (2018) {[}arXiv:1711.03390
{[}hep-th{]}{]}. 


\bibitem{Herdeiro:2018wub}C.~A.~R.~Herdeiro, E.~Radu, N.~Sanchis-Gual and J.~A.~Font,   ``Spontaneous Scalarization of Charged Black Holes,''   Phys.\ Rev.\ Lett.\  {\bf 121}, no. 10, 101102 (2018). [arXiv:1806.05190].  

%\bibitem{Cunha1904}P. V. Cunha, C. A. Herdeiro and E. Radu, Spontaneously Scalarized Kerr Black Holes in Extended Scalar-Tensor-Gauss-Bonnet Gravity, Phys. Rev. Lett. 123, no.1, 011101 (2019) {[}arXiv:1904.09997{[}gr-qc{]}{]}. 

%\bibitem{Dima:2020yac}A.~Dima, E.~Barausse, N.~Franchini and T.~P.~Sotiriou, ``Spin-induced black hole spontaneous scalarization,'' Phys. Rev. Lett. \textbf{125} (2020) no.23, 231101. [arXiv:2006.03095 [gr-qc]].

%\bibitem{Herdeiro2009}C. A. R. Herdeiro, E. Radu, H. O. Silva, T. P. Sotiriou and N. Yunes, Spin-induced scalarized black holes, Phys.Rev.Lett.126 (2021) 1, 011103. {[}arXiv:2009.03904 {[}gr-qc{]}{]}. 

%\bibitem{Berti2009}E. Berti, L. G. Collodel, B. Kleihaus and J. Kunz, Spin-induced black-hole scalarization in Einsteinscalar-Gauss-Bonnet theory, Phys.Rev.Lett. 126 (2021) 1, 011104. {[}arXiv:2009.03905 {[}gr-qc{]}{]}.

\bibitem{Blazquez-Salcedo:2020nhs}  J.~L.~Bl\'azquez-Salcedo, C.~A.~R.~Herdeiro, J.~Kunz, A.~M.~Pombo and E.~Radu, ``Einstein-Maxwell-scalar black holes: the hot, the cold and the bald,'' Phys. Lett. B \textbf{806}, 135493 (2020) %doi:10.1016/j.physletb.2020.135493 
[arXiv:2002.00963 [gr-qc]]. 

\bibitem{Blazquez-Salcedo:2020crd}   J.~L.~Bl\'azquez-Salcedo, S.~Kahlen and J.~Kunz, ``Critical solutions of scalarized black holes,'' Symmetry \textbf{12}, no.12, 2057 (2020) % doi:10.3390/sym12122057 
[arXiv:2011.01326 [gr-qc]].


\bibitem{Doneva:2021tvn}   D.~D.~Doneva and S.~S.~Yazadjiev,  ``Beyond the spontaneous scalarization: New fully nonlinear mechanism for the formation of scalarized black holes and its dynamical development,'' Phys. Rev. D \textbf{105}, no.4, L041502 (2022) %doi:10.1103/PhysRevD.105.L041502 
[arXiv:2107.01738 [gr-qc]].

\bibitem{Liu:2022fxy}
Y.~Liu, C.~Y.~Zhang, Q.~Chen, Z.~Cao, Y.~Tian and B.~Wang,
``The critical scalarization and descalarization of black holes in a generalized scalar-tensor theory,''
[arXiv:2208.07548 [gr-qc]].

\bibitem{Zhang:2021nnn}   C.~Y.~Zhang, Q.~Chen, Y.~Liu, W.~K.~Luo, Y.~Tian and B.~Wang,  ``Critical phenomena in dynamical scalarization of charged black hole,'' [arXiv:2112.07455 [gr-qc]].

%\bibitem{Gubser:2005ih} S.~S.~Gubser,``Phase transitions near black hole horizons,'' Class. Quant. Grav. \textbf{22}, 5121-5144 (2005)
%doi:10.1088/0264-9381/22/23/013 [arXiv:hep-th/0505189 [hep-th]].

%\bibitem{Stefanov:2007eq} I.~Z.~Stefanov, S.~S.~Yazadjiev and M.~D.~Todorov, ``Phases of 4D scalar-tensor black holes coupled to Born-Infeld nonlinear electrodynamics,'' Mod. Phys. Lett. A \textbf{23}, 2915-2931 (2008)
%doi:10.1142/S0217732308028351 [arXiv:0708.4141 [gr-qc]].

%\bibitem{Doneva:2010ke} D.~D.~Doneva, S.~S.~Yazadjiev, K.~D.~Kokkotas and I.~Z.~Stefanov, ``Quasi-normal modes, bifurcations and non-uniqueness of charged scalar-tensor black holes,'' Phys. Rev. D \textbf{82}, 064030 (2010)
%doi:10.1103/PhysRevD.82.064030 [arXiv:1007.1767 [gr-qc]].

%\bibitem{LuisBlazquez-Salcedo:2020rqp} J.~Luis Bl\'azquez-Salcedo, C.~A.~R.~Herdeiro, S.~Kahlen, J.~Kunz, A.~M.~Pombo and E.~Radu, ``Quasinormal modes of hot, cold and bald Einstein\textendash{}Maxwell-scalar black holes,'' Eur. Phys. J. C \textbf{81}, no.2, 155 (2021) %doi:10.1140/epjc/s10052-021-08952-w [arXiv:2008.11744 [gr-qc]].


\bibitem{Bizon:1998kq}   P.~Bizon and T.~Chmaj, ``Critical collapse of Skyrmions,'' Phys. Rev. D \textbf{58} (1998), 041501  [arXiv:gr-qc/9801012 [gr-qc]].

\bibitem{Choptuik:1996yg}   M.~W.~Choptuik, T.~Chmaj and P.~Bizon,  ``Critical behavior in gravitational collapse of a Yang-Mills field,'' Phys. Rev. Lett. \textbf{77} (1996), 424-427 [arXiv:gr-qc/9603051 [gr-qc]].


\bibitem{Bizon:2011gg}   P.~Bizon and A.~Rostworowski,  ``On weakly turbulent instability of anti-de Sitter space,'' Phys. Rev. Lett. \textbf{107} (2011), 031102  [arXiv:1104.3702 [gr-qc]].

\bibitem{Choptuik:1992jv}   M.~W.~Choptuik,  ``Universality and scaling in gravitational collapse of a massless scalar field,'' Phys. Rev. Lett. \textbf{70} (1993), 9-12  


\bibitem{Liebling:1996dx}
S.~L.~Liebling and M.~W.~Choptuik,
``Black hole criticality in the Brans-Dicke model,''
Phys. Rev. Lett. \textbf{77}, 1424-1427 (1996)
%doi:10.1103/PhysRevLett.77.1424
[arXiv:gr-qc/9606057 [gr-qc]].



\bibitem{Evans:1994pj} 
C.~R.~Evans and J.~S.~Coleman,
``Observation of critical phenomena and selfsimilarity in the gravitational collapse of radiation fluid,''
Phys. Rev. Lett. \textbf{72}, 1782-1785 (1994).
%doi:10.1103/PhysRevLett.72.1782 [arXiv:gr-qc/9402041 [gr-qc]].

\bibitem{Gundlach:2007gc}   C.~Gundlach and J.~M.~Martin-Garcia,  ``Critical phenomena in gravitational collapse,'' Living Rev. Rel. \textbf{10} (2007), 5  [arXiv:0711.4620 [gr-qc]].   


\bibitem{Bosch:2016vcp} 
  P.~Bosch, S.~R.~Green and L.~Lehner, ``Nonlinear Evolution and Final Fate of Charged Anti\textendash{}de Sitter Black Hole Superradiant Instability,'' Phys. Rev. Lett. \textbf{116}, no.14, 141102 (2016)  %doi:10.1103/PhysRevLett.116.141102 
[arXiv:1601.01384 [gr-qc]]. 

\bibitem{Chesler:2018txn}
P.~M.~Chesler and D.~A.~Lowe, ``Nonlinear Evolution of the AdS$_4$ Superradiant Instability,'' Phys. Rev. Lett. \textbf{122}, no.18, 181101 (2019) %doi:10.1103/PhysRevLett.122.181101 
[arXiv:1801.09711 [gr-qc]].


\bibitem{Berti:2009kk}
 E.~Berti, V.~Cardoso and A.~O.~Starinets, ``Quasinormal modes of black holes and black branes,'' Class. Quant. Grav. \textbf{26}, 163001 (2009) %doi:10.1088/0264-9381/26/16/163001 
[arXiv:0905.2975 [gr-qc]].

\bibitem{Brito:2015oca}
R.~Brito, V.~Cardoso and P.~Pani, ``Superradiance: New Frontiers in Black Hole Physics,'' Lect. Notes Phys. \textbf{906}, pp.1-237 (2015) %doi:10.1007/978-3-319-19000-6 
[arXiv:1501.06570 [gr-qc]].


%\bibitem{Fernandes:2019kmh}   P.~G.~S.~Fernandes, C.~A.~R.~Herdeiro, A.~M.~Pombo, E.~Radu and N.~Sanchis-Gual, ``Charged black holes with axionic-type couplings: Classes of solutions and dynamical scalarization,'' Phys. Rev. D \textbf{100}, no.8, 084045 (2019) %doi:10.1103/PhysRevD.100.084045 
%[arXiv:1908.00037].

%\bibitem{Ripley:2020vpk}    J.~L.~Ripley and F.~Pretorius, ``Dynamics of a $\mathbb Z_2$ symmetric EdGB gravity in spherical symmetry,'' Class. Quant. Grav. \textbf{37}, no.15, 155003 (2020) %doi:10.1088/1361-6382/ab9bbb 
%[arXiv:2005.05417 [gr-qc]].


%\bibitem{Kuan:2021lol}     H.~J.~Kuan, D.~D.~Doneva and S.~S.~Yazadjiev, ``Dynamical Formation of Scalarized Black Holes and Neutron Stars through Stellar Core Collapse,'' Phys. Rev. Lett. \textbf{127}, no.16, 161103 (2021) %doi:10.1103/PhysRevLett.127.161103 
%[arXiv:2103.11999 [gr-qc]].


%\bibitem{East:2021bqk}   W.~E.~East and J.~L.~Ripley, ``Dynamics of Spontaneous Black Hole Scalarization and Mergers in Einstein-Scalar-Gauss-Bonnet Gravity,'' Phys. Rev. Lett. \textbf{127}, no.10, 101102 (2021) %doi:10.1103/PhysRevLett.127.101102 
%[arXiv:2105.08571 [gr-qc]].

\bibitem{Silva:2020omi}
H.~O.~Silva, H.~Witek, M.~Elley and N.~Yunes,
``Dynamical Descalarization in Binary Black Hole Mergers,''
Phys. Rev. Lett. \textbf{127}, no.3, 031101 (2021)
%doi:10.1103/PhysRevLett.127.031101
[arXiv:2012.10436 [gr-qc]].

\bibitem{Doneva:2022byd}
D.~D.~Doneva, A.~Va\~n\'o-Vi\~nuales and S.~S.~Yazadjiev,
``Dynamical descalarization with a jump during black hole merger,''
[arXiv:2204.05333 [gr-qc]].

\bibitem{Corelli:2021ikv}
F.~Corelli, T.~Ikeda and P.~Pani,
``Challenging cosmic censorship in Einstein-Maxwell-scalar theory with numerically simulated gedanken experiments,''
Phys. Rev. D \textbf{104}, no.8, 084069 (2021)
%doi:10.1103/PhysRevD.104.084069
[arXiv:2108.08328 [gr-qc]].

\bibitem{Guo:2021zed} G.~Guo, P.~Wang, H.~Wu and H.~Yang, ``Scalarized Einstein\textendash{}Maxwell-scalar black holes in anti-de Sitter spacetime,'' Eur. Phys. J. C \textbf{81}, no.10, 864 (2021) %doi:10.1140/epjc/s10052-021-09614-7 
[arXiv:2102.04015 [gr-qc]].

\bibitem{Zhang:2021etr} C.~Y.~Zhang, P.~Liu, Y.~Liu, C.~Niu and B.~Wang,  ``Dynamical charged black hole spontaneous scalarization in anti\textendash{}de Sitter spacetimes,'' Phys. Rev. D \textbf{104} (2021) no.8, 084089   [arXiv:2103.13599 [gr-qc]].


%\bibitem{Santos-Olivan:2015yok}  D.~Santos-Oliv\'an and C.~F.~Sopuerta,  ``New Features of Gravitational Collapse in Anti\textendash{}de Sitter Spacetimes,'' Phys. Rev. Lett. \textbf{116}, no.4, 041101 (2016) [arXiv:1511.04344 [gr-qc]].



\bibitem{Chesler:2013lia}     P.~M.~Chesler and L.~G.~Yaffe, ``Numerical solution of gravitational dynamics in asymptotically anti-de Sitter spacetimes,'' JHEP \textbf{07}, 086 (2014)  % doi:10.1007/JHEP07(2014)086 
[arXiv:1309.1439 [hep-th]].

%\bibitem{Arnowitt:1959ah}
%R.~L.~Arnowitt, S.~Deser and C.~W.~Misner,
%``Dynamical Structure and Definition of Energy in General Relativity,''
%Phys. Rev. \textbf{116}, 1322-1330 (1959)
%doi:10.1103/PhysRev.116.1322

\bibitem{Zhang:2021ybj}   C.~Y.~Zhang, P.~Liu, Y.~Liu, C.~Niu and B.~Wang, ``Dynamical scalarization in Einstein-Maxwell-dilaton theory,'' Phys. Rev. D \textbf{105}, no.2, 024073 (2022) %doi:10.1103/PhysRevD.105.024073 
[arXiv:2111.10744 [gr-qc]].

\bibitem{Zhang:2021edm}    C.~Y.~Zhang, P.~Liu, Y.~Liu, C.~Niu and B.~Wang, ``Evolution of anti\textendash{}de Sitter black holes in Einstein-Maxwell-dilaton theory,'' Phys. Rev. D \textbf{105}, no.2, 024010 (2022) %doi:10.1103/PhysRevD.105.024010
[arXiv:2104.07281].

\bibitem{Luo:2022roz}
W.~K.~Luo, C.~Y.~Zhang, P.~Liu, C.~Niu and B.~Wang,
``Dynamical spontaneous scalarization in Einstein-Maxwell-scalar models in anti-de Sitter spacetime,''
[arXiv:2206.05690 [gr-qc]].

\bibitem{Herdeiro:2020xmb}
 C.~A.~R.~Herdeiro and E.~Radu, ``Spherical electro-vacuum black holes with resonant, scalar $Q$-hair,'' Eur. Phys. J. C \textbf{80}, no.5, 390 (2020) %doi:10.1140/epjc/s10052-020-7976-9 
[arXiv:2004.00336 [gr-qc]].

\bibitem{Hong:2020miv}
J.~P.~Hong, M.~Suzuki and M.~Yamada, ``Spherically Symmetric Scalar Hair for Charged Black Holes,'' Phys. Rev. Lett. \textbf{125}, no.11, 111104 (2020) %doi:10.1103/PhysRevLett.125.111104 
[arXiv:2004.03148 [gr-qc]].

\bibitem{Emparan:2001wn}
R.~Emparan and H.~S.~Reall, ``A Rotating black ring solution in five-dimensions,'' Phys. Rev. Lett. \textbf{88}, 101101 (2002) %doi:10.1103/PhysRevLett.88.101101 
[arXiv:hep-th/0110260 [hep-th]].

\bibitem{Emparan:2007wm}
R.~Emparan, T.~Harmark, V.~Niarchos, N.~A.~Obers and M.~J.~Rodriguez, ``The Phase Structure of Higher-Dimensional Black Rings and Black Holes,'' JHEP \textbf{10}, 110 (2007) %doi:10.1088/1126-6708/2007/10/110 
[arXiv:0708.2181 [hep-th]].



\bibitem{Gubser:2008ny} 
 S.~S.~Gubser and A.~Nellore, ``Mimicking the QCD equation of state with a dual black hole,'' Phys. Rev. D \textbf{78}, 086007 (2008) %doi:10.1103/PhysRevD.78.086007 
[arXiv:0804.0434 [hep-th]].

\bibitem{Janik:2017ykj} 
R.~A.~Janik, J.~Jankowski and H.~Soltanpanahi, ``Real-Time dynamics and phase separation in a holographic first order phase transition,'' Phys. Rev. Lett. \textbf{119}, no.26, 261601 (2017) %doi:10.1103/PhysRevLett.119.261601 
[arXiv:1704.05387 [hep-th]].

\bibitem{Attems:2019yqn} 
M.~Attems, Y.~Bea, J.~Casalderrey-Solana, D.~Mateos and M.~Zilh\~ao, ``Dynamics of Phase Separation from Holography,'' JHEP \textbf{01}, 106 (2020) %doi:10.1007/JHEP01(2020)106 
[arXiv:1905.12544 [hep-th]].

\bibitem{Most:2018eaw} 
 E.~R.~Most, L.~J.~Papenfort, V.~Dexheimer, M.~Hanauske, S.~Schramm, H.~St\"ocker and L.~Rezzolla, ``Signatures of quark-hadron phase transitions in general-relativistic neutron-star mergers,'' Phys. Rev. Lett. \textbf{122}, no.6, 061101 (2019) %doi:10.1103/PhysRevLett.122.061101
[arXiv:1807.03684 [astro-ph.HE]].

\bibitem{Zha:2020gjw} 
 S.~Zha, E.~P.~O'Connor, M.~c.~Chu, L.~M.~Lin and S.~M.~Couch, ``Gravitational-Wave Signature of a First-Order Quantum Chromodynamics Phase Transition in Core-Collapse Supernovae,'' Phys. Rev. Lett. \textbf{125}, no.5, 051102 (2020) 
 %[erratum: Phys. Rev. Lett. \textbf{127}, no.21, 219901 (2021)] %doi:10.1103/PhysRevLett.127.219901 
[arXiv:2007.04716 [astro-ph.HE]].

\end{thebibliography}
\end{document}